\documentclass{aastex}
\usepackage{emulateapj5}
\begin{document}

\newcommand{\hi}{$h^{-1}$~}
\newcommand{\kms}{~km~s$^{-1}$}
\newcommand{\logh}{$+5\log h$}

\title{Infall, the Butcher-Oemler Effect, and the Descendants of Blue
  Cluster Galaxies at $z\sim0.6^{1,2}$}

\author{Kim-Vy H. Tran$^3$, Pieter van Dokkum$^4$, Garth D. Illingworth$^5$, \\
Daniel Kelson$^6$, Anthony Gonzalez$^{7,8}$, \& Marijn Franx$^9$ }

\footnotetext[1]{Based on observations with the NASA/ESA Hubble Space
  Telescope, obtained at the Space Telescope Science Institute, which
  is operated by the Association of Universities for Research in
  Astronomy, Inc., under NASA contract NAS 5-26555.}
\footnotetext[2]{Based on observations obtained at the W. M. Keck
  Observatory, which is operated jointly by the California Institute of
  Technology and the University of California.}
\footnotetext[3]{Institute for Astronomy, ETH Z\"urich, CH-8093
  Z\"urich, Switzerland, vy@phys.ethz.ch}
\footnotetext[4]{Department of Astronomy, Yale University, New Haven,
  CT 06520-8101}
\footnotetext[5]{University of California Observatories/Lick
    Observatory, University of California, Santa Cruz, CA 95064}
\footnotetext[6]{Observatories of the Carnegie Institution of
  Washington, 813 Santa Barbara Street, Pasadena, CA, 91101}
\footnotetext[7]{NSF Astronomy and Astrophysics Postdoctoral Fellow}
\footnotetext[8]{  Department of Astronomy, University of Florida,
  Gainesville, FL 32611}
\footnotetext[9]{Leiden Observatory, P.O. Box 9513, 2300 RA Leiden,
    The Netherlands}

\setcounter{footnote}{9}

\begin{abstract}
  
  Using wide-field HST/WFPC2 imaging and extensive Keck/LRIS
  spectroscopy, we present a detailed study of the galaxy populations
  in MS~2053--04, a massive, X-ray luminous cluster at
  $z=0.5866\pm0.0011$.  Analysis of 149 confirmed cluster members shows
  that MS2053 is composed of two structures that are gravitationally
  bound to each other; their respective velocity dispersions are
  $865\pm71$\kms~(113 members) and $282\pm51$\kms~(36 members).
  MS2053's total dynamical mass is $1.2\times10^{15}M_{\odot}$.  MS2053
  is a classic Butcher-Oemler cluster with a high fraction of blue
  members ($24\pm5$\%) and an even higher fraction of star-forming
  members ($44\pm7$\%), as determined from their [OII]$\lambda3727$
  emission.  The number fraction of blue/star-forming galaxies is much
  higher in the infalling structure than in the main cluster. This
  result is the most direct evidence to date that the Butcher-Oemler
  effect is linked to galaxy infall.  In terms of their colors,
  luminosities, estimated internal velocity dispersions, and
  [OII]$\lambda3727$ equivalent widths, the infalling galaxies are
  indistinguishable from the field population.  MS2053's deficit of S0
  galaxies combined with its overabundance of blue spirals implies that
  many of these late-types will evolve into S0 members.  The properties
  of the blue cluster members in both the main cluster and infalling
  structure indicate they will evolve into {\it low mass}, $L<L^{\ast}$
  galaxies with extended star formation histories like that of low mass
  S0's in Coma.  Our observations show that most of MS2053's blue
  cluster members, and ultimately most of its low mass S0's, originate
  in the field.  Finally, we measure the redshift of the giant arc in
  MS2053 to be $z=3.1462$; this object is one in only a small set of
  known strongly lensed galaxies at $z>3$.

\end{abstract}


\section{Introduction}

The seminal papers by \citet{butcher:78,butcher:84} gave rise to an
entire generation of studies on the evolution of galaxy populations in
clusters.  Their observations of an increasing number of blue cluster
members with increasing redshift, christened the Butcher-Oemler (B-O)
effect, have since been confirmed with photometric
\citep{couch:94,dressler:94,rakos:95,lubin:96,dressler:97,couch:98,margoniner:00}
and spectroscopic studies
\citep{lavery:86,couch:87,newberry:88,lavery:92,dressler:92,abraham:96b,caldwell:97,ellingson:01,tran:03b}.
However, the origins of the B-O effect have yet to be fully understood.

Several early studies suggested the B-O effect was due to galaxy-galaxy
interactions \citep{lavery:88,lavery:92,dressler:94,couch:94}.  More
recently, galaxy infall has become a favored mechanism for increasing
the blue fraction in clusters, especially at intermediate redshifts
\citep{kauffmann:95,couch:98,vandokkum:98a,ellingson:01,fairley:02}.  A
panoramic imaging study of A851 ($z=0.41$) by \citet{kodama:01b}
suggests that the transition from predominantly blue to red galaxy
colors occurs most often in the subclumps outside the cluster core.
However, the observations needed to understand the physical mechanisms
driving the evolution of cluster galaxies, $e.g.$ redshifts,
morphologies, and spectral types, are difficult to obtain.  Although
substructure in massive clusters is quite common and is attributed to
the continued accretion of galaxies, the evidence linking galaxy infall
to the B-O effect, $e.g.$ the spatial and kinematic segregation of
early and late-type members \citep{dressler:80,fisher:98,biviano:02},
remains circumstantial.

Knowing whether the B-O effect is due to galaxy infall is particularly
important if we are to explain the unexpected deficit of S0 galaxies
observed at intermediate redshifts
\citep{dressler:97,fasano:00,lubin:02}.  These studies suggest that the
S0's in nearby clusters form out of the excess of spirals observed at
higher redshifts, $i.e.$ the B-O galaxies \citep[although for an
alternative explanation, see][]{andreon:98}.  If S0's form from
infalling spirals, then this would mean that most S0's in nearby
clusters originated in the field.

To assess how galaxy infall, the Butcher-Oemler effect, and the
possible progenitors of S0 members are related, we present a detailed
study of MS~2053--04, a massive, X-ray luminous galaxy cluster
\citep{stocke:91,luppino:92} at $z\sim0.6$.  With extensive Keck/LRIS
spectroscopy, we identify 149 members and measure MS2053's velocity
dispersion and mass.  Combining our spectroscopic survey with
wide-field HST/WFPC2 imaging, we then examine the colors, luminosities,
morphologies, estimated internal velocity dispersions, and
[OII]$\lambda3727$ equivalent widths of the confirmed cluster members.
We also test whether the current star formation in B-O galaxies at
$z\sim0.6$ is consistent with trends observed in clusters at $z<0.2$,
$e.g.$ that cluster S0's tend to have younger luminosity-weighted ages
than the ellipticals \citep{poggianti:01b,smail:01}.  For comparison to
the field population, we utilize a field sample with redshifts
comparable to MS2053.

In \S2 and \S3, we describe our imaging and spectroscopic observations
as well as how physical properties such as colors and spectral types
were determined. In \S4, we briefly summarize our sample of field
galaxies.  In \S5, we present MS2053's redshift distribution, velocity
dispersion, and mass as well as a short discussion on substructure.  In
\S6, we examine MS2053's early-type galaxy population.  We determine
MS2053's fraction of blue members and show how it is related to galaxy
infall in \S7.  The spectroscopic Butcher-Oemler fraction is presented
in \S8 along with a summary of the X-ray observations from Chandra.  We
dicuss how the blue members in MS2053 are related to S0 galaxies in
lower redshift clusters in \S9, and present our conclusions in \S10.

Unless otherwise noted, we use $\Omega_M=0.3, \Omega_{\Lambda}=0.7$,
and $H_0=100h$\kms~Mpc$^{-1}$.  At MS2053's redshift, this corresponds
to a projected scale of $4.6$\hi~kpc/arcsec.

\section{Imaging}

\subsection{Ground-Based Observations}

Our early observations of the MS2053 field included a 200s $R$ and 240s
$I$ image ($1.1''$ seeing) taken at Keck centered on the Brightest
Cluster Galaxy (BCG; \#1667).  Objects were detected and magnitudes
determined with FOCAS \citep{valdes:82}.  The $R$ and $I$ images were
used primarily to select spectroscopic targets for observing runs from
August 1995 up to August 1997.

\subsection{HST/WFPC2 Observations}

\begin{figure}
\epsscale{0.5}
\plotone{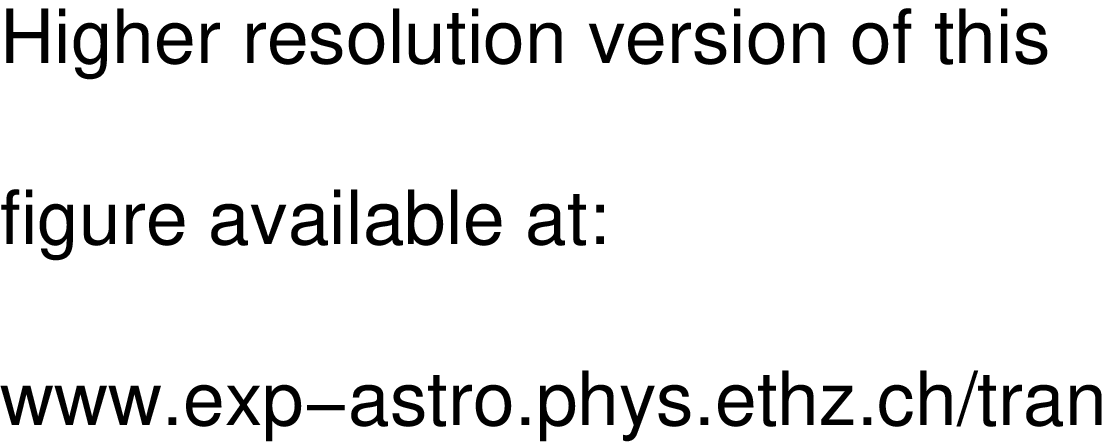}
\caption{HST/WFPC2 false-color mosaic of the MS2053 field.  The
  $\sim4'\times7'$ mosaic is made of six overlapping pointings taken in
  F606W and F814W; the projected scale at $z=0.59$ is 4.6\hi~kpc/arcsec
  ($\Omega_M=0.3, \Omega_{\Lambda}=0.7$, and
  $H_0=100h$\kms~Mpc$^{-1}$).  Although MS2053 has 149
  spectroscopically confirmed members, a high velocity dispersion
  ($\sigma=865$\kms), and a high X-ray temperature
  \citep[5.2~keV;][]{vikhlinin:02}, it is not a visually striking
  cluster.
\label{mosaic}}
\end{figure}

MS2053 was imaged by HST/WFPC2 in October 1998.  The observations
consisted of six overlapping pointings taken in both F606W and F814W
filters and covered an area $\sim4'\times7'$ (Fig.~\ref{mosaic}).  The
total integration time at each pointing was 3200s and 3300s for F606W
and F814W respectively; the image reduction is detailed in
\citet{hoekstra:02}.

Total magnitudes were measured using SExtractor \citep{bertin:96}, and
$(B-V)_z$ colors measured within a $3''$ diameter aperture.  Following
the method outlined in \citet{vandokkum:96}, we transform magnitudes
in the WFPC2 filter system to redshifted Johnson magnitudes using:

\begin{equation}
B_z=F814W + 0.354(F606W-F814W) +0.923
\end{equation}
\begin{equation}
V_z=F814W-0.173(F606W-F814W)+0.959
\end{equation}

\noindent where the constants  were calculated for an E/S0 galaxy
spectral energy distribution \citep{pence:76} redshifted to $z=0.59$.
Apparent magnitude is converted to an absolute magnitude using a
distance modulus adjusted for passive evolution (($m-M=41.2$); here we
have accounted for simple fading, as determined from the Fundamental
Plane \citep[$\Delta\log(M/L_B)\propto-0.40z$;][]{vandokkum:98b}.  We
also correct for Galactic reddening as $E(B-V)=0.084$ mag in the MS2053
field \citep{schlegel:98}.

\subsection{Morphologies and Structural Parameters}

All galaxies on the WFPC2 imaging brighter than $m_{814}=22$ were
visually classified by M. Franx, P. van Dokkum, and D. Fabricant using
the method described in \citet{fabricant:00}.  The morphological types
of \{E, E/S0, S0, S0/a, Sa, Sb, Sc, Sd\} were assigned values of
$\{-5, -4, -2, 0, 1, 3, 5, 7\}$; intermediate values $\{-3, -1, 2\}$
were also used.  In our analysis, the galaxies are split into five main
categories: E/S0 galaxies have $-5\leq T\leq-1$, S0/a galaxies $T=0$,
Sa galaxies $T=1$, spirals $T\geq2$, and mergers $T=99$.

We complement the visual classifications with structural parameters
measured using GIM2D \citep{simard:02}.  For each cluster and field
galaxy with HST/WFPC2 imaging, we fit surface brightness models using a
pure de Vaucouleurs profile.  The half-light radii and surface
brightnesses were then used with the galaxy colors to estimate internal
velocity dispersions (see \S\ref{sigma}).  For a more extensive
discussion on analyzing cluster galaxies at intermediate redshifts
using GIM2D, see \citet{tran:03a}.

\section{Spectroscopy}

\subsection{Observations and Reductions}

Using LRIS \citep{oke:95} on Keck, a spectroscopic survey of the MS2053
field was conducted from August 1995 to July 2001 over 7 observing
runs.  During this period, spectroscopic targets were selected from
three object catalogs.  The first two catalogs were made from the Keck
$R$ and $I$ images.  Spectra taken during the August 1995 to August
1997 observing runs were selected from these Keck catalogs.  A third
object catalog was created later from the HST/WFPC2 mosaic (taken in
October 1998).  Spectra obtained after August 1997 were selected from
the HST catalog.  In all cases, the spectroscopic targets were selected
by magnitude ($m_{814}\leq22$) and not morphology.

A total of 18 multi-slit masks containing 669 targets were observed.
The multi-slit masks included both masks designed to measure redshifts
($t_{\rm exp}\sim2000$ sec) and masks to measure internal velocity
dispersions ($t_{\rm exp}\geq7500$ sec) with redshift fillers.  The
slits were typically $1''$ wide and the seeing on the observing runs
was $\leq1.''2$.  Depending on the grating used, the spectral
resolution (FWHM) ranged from $9-13$\AA~with higher spectral resolution
($5-6$\AA) used for the dispersion masks.

IRAF\footnote{IRAF is distributed by the National Optical Astronomy
  Observatories, which are operated by the Association of Universities
  for Research in Astronomy, Inc., under cooperative agreement with the
  National Science Foundation.} routines and custom software provided
by L. Simard and D. Kelson \citep{kelson:98} were used to reduce the
spectra; a detailed explanation of the reduction pipeline can be found
in \citet{tran:99}.  The spectra were corrected for the telluric
atmospheric A and B bands by using the spectrum of a bright blue star
included on all the masks.  The observed wavelength coverage of the 669
targets depended on the grating and slitlet position, but was typically
$6000-9500$\AA.  For most cluster members, this includes the
[OII]~$\lambda3727$ doublet and 4000\AA~break.

\subsection{Measuring Redshifts}

The IRAF routine XCSAO \citep{kurtz:92} was used to measure absorption
line redshifts and their errors.  Four template galaxy spectra were
used: NGC7331 (morphological type SA(s)b), NGC4889 (E4), NGC2276
(SAB(rs)c), and an ``E+A'' galaxy.  The cross-correlation wavelength
range for the cluster members was approximately $3750-4500$\AA~in the
restframe.  Emission line redshifts were determined from measuring the
central wavelengths of the appropriate lines, $e.g.$ the
[OII]$\lambda3727$ doublet, H$\beta$, and/or [OIII]$\lambda5007$.  All
redshifts were inspected by eye.  No systematic difference was found
for redshifts determined from both absorption and emission lines.

The final redshift catalog has 484 redshifts: 44 stars, 157 cluster
members, and 283 field galaxies.  Each redshift was given a quality
flag where $Q=3, 2, \&~1$ corresponded to definite, probable, and maybe
(single emission line).  In our analysis, we consider only galaxies
with a redshift quality flag of $Q=3$; this reduces the total cluster
sample to 149 members.  The average redshift error is $\sim30$\kms.

\subsection{Completeness and Selection Effects}

To determine if our redshift survey of the MS2053 field is influenced
by selection effects, we consider the possibility of magnitude bias due
to it being inherently more difficult to measure redshifts of fainter
objects, and color bias due to sparse sampling.  In the following
discussion, we consider only the objects that fall on the WFPC2 mosaic.

\begin{figure}
\epsscale{0.8}
\plotone{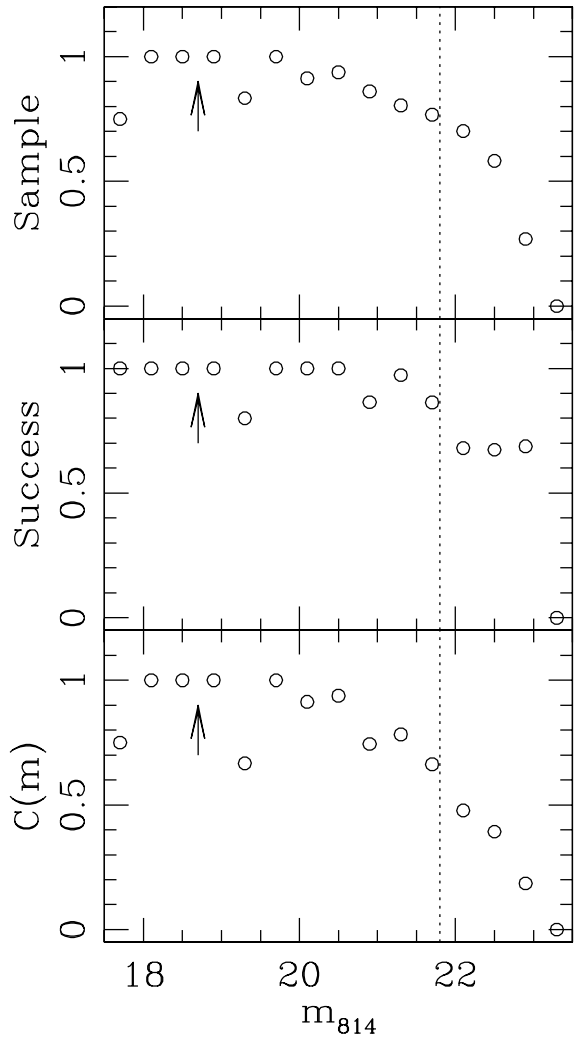}
\caption{Our Keck/LRIS spectroscopic survey of the MS2053 field is
  magnitude limited such that incompleteness at the faint end is due to
  sparse sampling and not the inability to measure redshifts of
  targeted galaxies.  {\it Top:} The sampling rate, defined as the
  number of spectroscopic targets divided by the number of galaxies in
  the HST/WFPC2 photometric catalog, is shown as a function of
  magnitude (bin size is 0.4 mags); only objects on the WFPC2 mosaic
  are considered.  The BCG's (\#1667) magnitude ($m_{814}=18.7$) is
  indicated by the arrow.  {\it Middle:} The number of acquired
  redshifts divided by the number of targets.  At $m_{814}=22$, the
  success rate is $\sim70$\%.  {\it Bottom:} The completeness $C(m)$ is
  the product of the sampling and success rates.  The dotted vertical
  line denotes the approximate $m_{814}$ value corresponding to
  $M_{Be}=-18$\logh, the magnitude cut-off used in our analysis.
\label{rates}}
\end{figure}

By comparing the number of galaxies in the WFPC2 photometric catalog to
the number of spectroscopic targets and acquired redshifts, we
investigate the completeness of our survey using the method outlined in
\citet{yee:96} and \citet{vandokkum:00}.  Figure~\ref{rates} shows the
sampling rate, success rate, and completeness of our sample as a
function of $m_{814}$.  The sampling rate, defined as the number of
spectroscopic targets divided by the number of galaxies in the HST
catalog, is $\sim70$\% at $m_{814}=22$ ($M_{Be}\sim-17.7$\logh).  The
success rate, defined as the number of acquired redshifts divided by
the number of targets, is also high ($\sim70$\%) at $m_{814}=22$.  Our
incompleteness at the faint end is due to sparse sampling and not the
inability to measure redshifts of targeted galaxies.

\begin{figure}
\epsscale{0.6}
\plotone{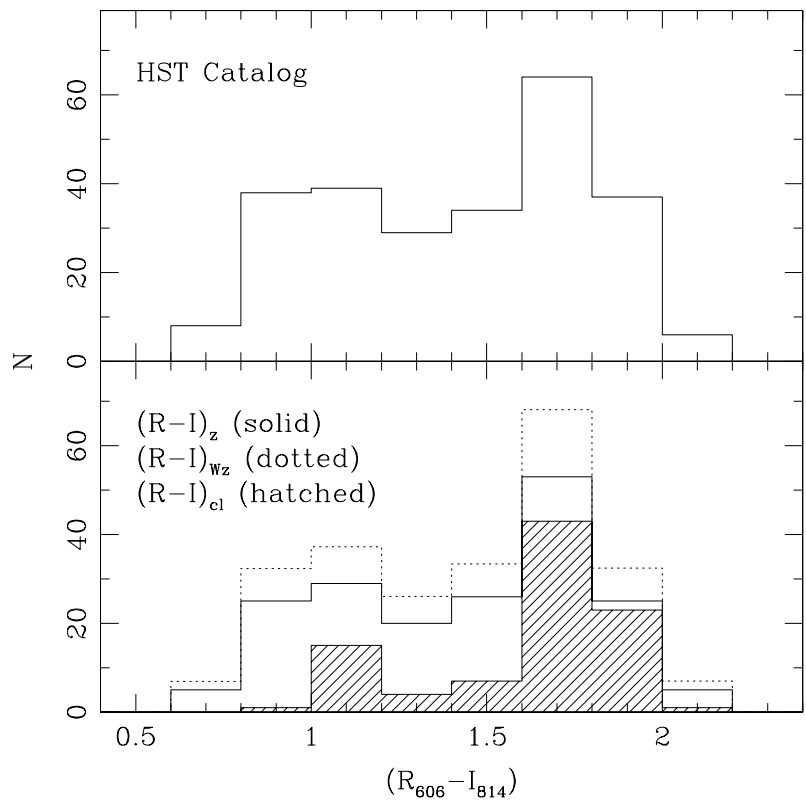}
\caption{Our spectroscopic sample is not biased against faint blue
  galaxies.  {\it Top:} The color distribution $(R_{606}-I_{814})$ of
  all galaxies in the WFPC2 catalog with $20<m_{814}\leq22$ (378); the
  bin size is $\Delta(R-I)=0.2$.  {\it Bottom:} The color distribution
  of all galaxies with measured redshifts (solid line; $(R-I)_z$) as
  well as cluster members (hatched; $(R-I)_{cl}$) in the same
  magnitude range.  We also include the weighted color distribution
  $(R-I)_{Wz}$ (dotted) of the redshift sample where each galaxy is
  weighted by the inverse of the magnitude selection function $C(m)$
  to correct for sparse sampling.  The color distributions of the
  photometric and spectroscopic samples are indistinguishable using
  the K-S test.
\label{ri_hist}}
\end{figure}

We then compare the colors of galaxies with measured redshifts to all
galaxies in the WFPC2 catalog to determine if the spectroscopic sample
is biased against faint blue galaxies because of sparse sampling.  We
focus on objects with $20<m_{814}\leq22$ and denote
$(R_{606}-I_{814})$ as $(R-I)$.  Figure~\ref{ri_hist} compares the
color distribution of all galaxies in the WFPC2 catalog to that of the
spectroscopic sample and cluster members.  Note the similarity in the
color distributions of the photometric sample to that of the
spectroscopic sample; a Kolmogorov-Smirnov test \citep{press:92} finds
the two distributions are indistinguishable, $i.e.$, our redshift
sample is not biased against faint blue galaxies.  As an additional
visual check, we include the weighted color distribution of the
spectroscopic sample $(R-I)_{Wz}$ determined by weighting each galaxy
by the inverse of the magnitude selection function $C(m)$; this is
also quite similar to the full photometric sample.

\subsection{Spectral Types}  

Using the same spectral bandpasses as \citet{fisher:98} and
\citet{tran:03b,tran:04a}, we measure [OII]$\lambda3727$, H$\delta$,
H$\gamma$, and H$\beta$ equivalent widths (EW) for the 149 cluster
members and 38 field galaxies (see \S4).  We separate the galaxies into
three spectral types to quantify their recent and ongoing star
formation: 1) absorption line galaxies with no significant [OII]
emission ([OII]$<5$\AA) and no strong Balmer absorption
[(H$\delta+$H$\gamma)/2>-4$~\AA]; 2) emission line galaxies with strong
[OII] emission ($\geq5$\AA); and 3) post-starburst (``E+A'') galaxies
that have weak or no [OII]$\lambda3727$ emission ($\leq5$~\AA) and
strong Balmer absorption [(H$\delta+$H$\gamma)/2\leq-4$~\AA].

\subsection{Internal Velocity Dispersions}\label{sigma}

A significant part of our spectroscopic observations with Keck/LRIS
were devoted to measuring internal velocity dispersions for the
brightest cluster members in MS2053.  The reduction and analysis of
the measured internal velocity dispersions ($\sigma$) for 33 cluster
galaxies ($-21.0\leq M_{Be}-5\log h \leq -18.1$) are described in
\citet{wuyts:04}.  

To estimate internal velocity dispersions ($\sigma_{est}$) for the rest
of our cluster sample, we use the method outlined in \citet{kelson:00c}
and \citet{tran:03b}.  To summarize, we evolve the galaxies until they
lie on the color magnitude relation defined by a passively evolving
galaxy population, $i.e.$ the early-types in MS2053, and then use the
Fundamental Plane to estimate $\sigma$.  Because this method requires
accurate colors and structural parameters, we can estimate velocity
dispersions only for galaxies that have WFPC2 imaging.  We measure the
necessary structural parameters by fitting pure de Vaucouleurs profiles
to the galaxies (see \S2.3).

\subsection{Redshift of the Gravitational Arc}

\begin{figure}
\epsscale{0.5}
\rotate
\plotone{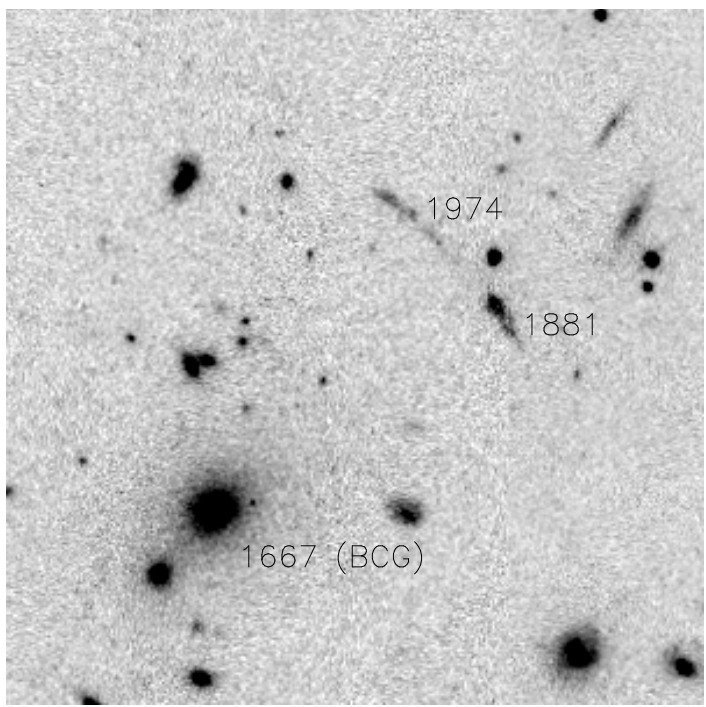}
\caption{HST/WFPC2 F814W image ($30''\times 30''$) of the two giant
  gravitational arcs in MS2053; north and east are approximately up
  and to the left.  The BCG is at $(20^h 56' 21.3'', -4^{\circ} 37'
  50.7'')_{2000}$ and the brighter of the two arcs (\#1881) is at
  $(20^h 56' 20.8'', -4^{\circ} 37' 38.1'')_{2000}$.
\label{hst_arc}}
\end{figure}

\begin{figure}
\epsscale{0.6}
\plotone{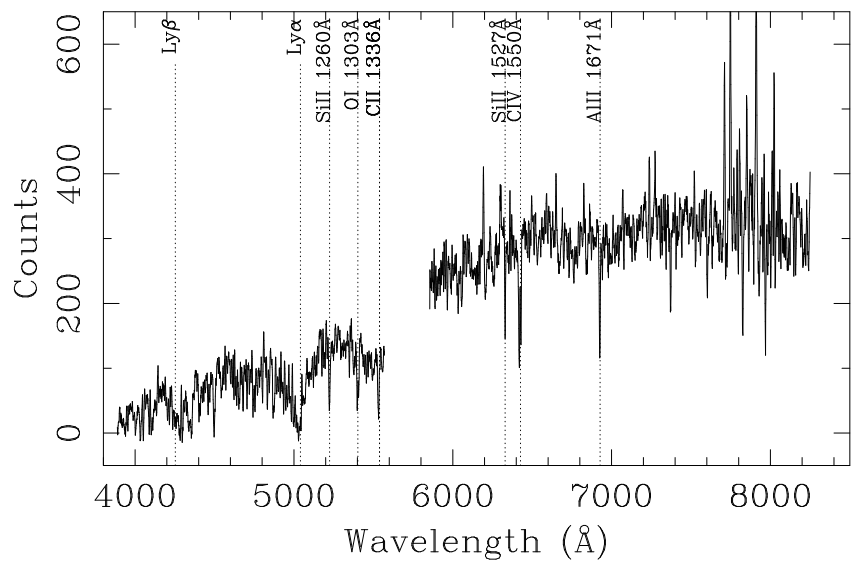}
\caption{Keck/LRIS spectrum of the giant gravitational arc \#1881 in
MS2053.  The discontinuity at $\sim5800$\AA~corresponds to the break
between the blue and red sides of the LRIS observations.  The source
redshift is at $z=3.1462$; this object is only one of a handful of
known strongly lensed galaxies at $z>3$.
\label{lens}}
\end{figure}

MS2053 lenses a background galaxy that appears as two giant tangential
arcs \citep[\#1881 \& \#1974 in Fig.~\ref{hst_arc}]{luppino:92}.  After
a heroic effort that spanned virtually all of our Keck/LRIS observing
runs for MS2053, we were able to measure in July 2001 a redshift for
galaxy \#1881 [$(20^h 56' 20.8'', -4^{\circ} 37' 38.1'')_{2000}$].
With a total integration time of 3.5 hours, we determine the source
redshift to be $z=3.1462$; the 1D spectrum is shown in Fig.~\ref{lens}.
This object is one of only a small set of known strongly lensed galaxies
at $z>3$.

\section{Field Sample}

In our analysis, we also include a comparison to field galaxies at the
same approximate redshift as MS2053.  The field galaxies are drawn from
an extensive spectroscopic survey completed by our group of four
different fields, of which MS2053 was one.  Our unique dataset contains
an unsually large number of spectroscopically confirmed field galaxies
($\sim800$; $0.05<z<3.15$) gathered over a total area of
$\sim200\Box'$; HST/WFPC2 mosaics in F606W and F814W were also obtained
in these fields.  The observational details of the fields and their
characteristics are summarized in \citet{tran:03b,tran:04a}.

We select field galaxies within the redshift range $0.54<z<0.64$ whose
wavelength coverage included [OII]$\lambda3727$, and that were imaged
by WFPC2.  These selection criteria enable direct comparison between
the field population and the cluster members since both can be analyzed
in the exact same manner, $i.e.$ spectral indices, $(B-V)_z$ colors,
rest-frame magnitudes, morphological types, and structural parameters.
Our field sample of 38 galaxies spans comparable ranges in luminosity
($-16.3<M_{Be}-5\log h<-20.9$) and morphology ($-5\leq T\leq15$) as
the cluster members.  We also estimate internal velocity dispersions
for the field galaxies using the method described in \S\ref{sigma}
\citep[see][]{tran:03b,tran:04a}.

\section{Cluster Characteristics}

\subsection{MS2053's Redshift and Velocity Dispersion}

\begin{figure}
\epsscale{0.8}
\plotone{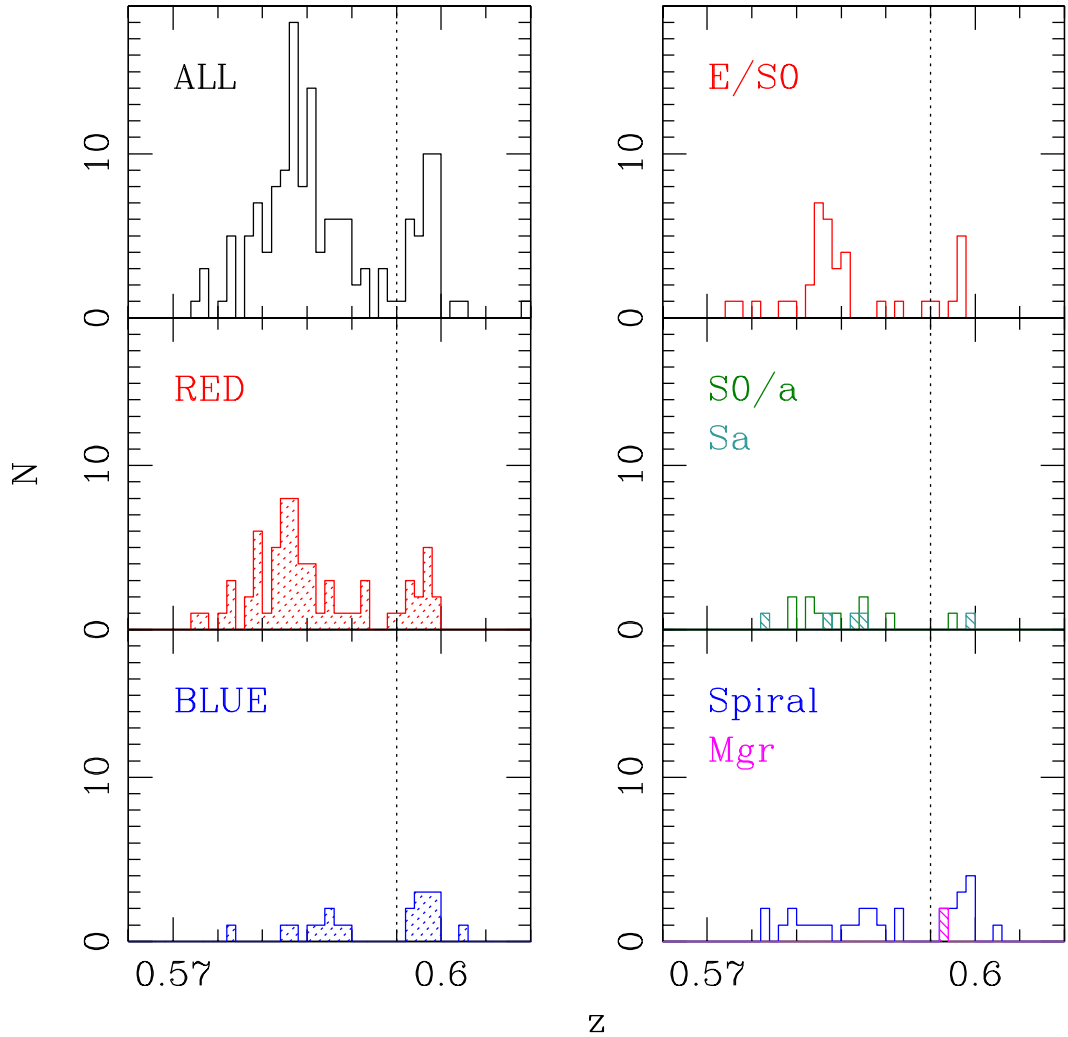}
\caption{Redshift histogram of MS2053 based on our 149 confirmed
  cluster members.  We find two pronouced redshift peaks, one at
  $z=0.5840$ (MS2053-A) and a smaller one at $z=0.5983$ (MS2053-B); the
  vertical dotted line indicates the redshift that we used to separate
  the two components.  In the other five panels, we consider only the
  cluster members on the HST/WFPC2 mosaic that are brighter than
  $M_{Be}=-18$\logh~(89).  The bottom left two panels show the blue
  ($\Delta (B-V)_z\leq-0.2$) and red ($\Delta (B-V)_z>-0.2$) members,
  while the three right panels show the members separated by
  morphological type.  More than half of the blue/late-type cluster
  members are in MS2053-B.
\label{zhist}}
\end{figure}

From our redshift survey of the MS2053 field, we identify 149 cluster
members and show their redshift distribution in Fig.~\ref{zhist}.  The
cluster members show a striking bimodal distribution with a main peak
centered at $z=0.5840\pm0.0005$ and a smaller peak at
$z=0.5982\pm0.0003$.  Guided by the shapes of the two peaks, we define
members with $0.57\leq z<0.595$ to be in the main cluster (hereafter
referred to as MS2053-A) and members at $0.595\leq z\leq0.605$ to be in
the group/sheet (MS2053-B).  MS2053-A has 113 galaxies and MS2053-B 36
galaxies.

The velocity dispersions of MS2053-A and MS2053-B are
$865\pm71$\kms~and $282\pm51$\kms~respectively.  The average redshifts,
dispersions, and their corresponding errors are calculated using the
biweight and jacknife methods \citep{beers:90}, and they are listed in
Table~\ref{stats}.  The velocity dispersion of MS2053-A is in
remarkable agreement with that estimated from weak-lensing
\citep[$886\pm139$\kms;][]{hoekstra:02}.  If MS2053's bimodal redshift
distribution had not been recognized, the cluster's spectroscopic
velocity dispersion would be significantly overestimated
($\sim1500$\kms).

\subsection{Virial Mass}

To estimate the virial masses of MS2053-A and MS2053-B, we follow
\citet{ramella:89} and first determine the virial radii using:

\begin{equation}
R_V = \frac{\pi\bar{V}}{H_0} 
\sin \left\{ \frac{1}{2} \left[ \frac{N_m(N_m-1)}{2} \left( \sum_{i}
      \sum_{j>i} \theta_{ij}^{-1} \right)^{-1} \right] \right\}
\end{equation}

\noindent where $\bar{V}$ is the mean velocity of the cluster, $N_m$ is the
number of members, and $\theta_{ij}$ the angular separation between the
$i^{th}$ and $j^{th}$ members.  With the virial radius, we then
estimate the mass using

\begin{equation}
M_V = \frac{6\sigma_{1D}^2 R_V}{G}
\end{equation}

\noindent where $\sigma_{1D}$ is the line of sight velocity dispersion
and $G$ the gravitational constant.  We consider MS2053-A and MS2053-B
separately and measure their virial radii to be 1.07 and 1.13\hi~Mpc,
respectively.  MS2053-A's mass is $\sim1.1\times10^{15}M_{\odot}$ while
MS2053-B is only $\sim1.1\times10^{14}M_{\odot}$ (Table~\ref{stats}).

Combining MS2053-A's velocity dispersion with the X-ray temperature
measured from Chandra for MS2053 \citep[$T_x=5.2\pm0.7$
keV;][]{vikhlinin:02}, we find MS2053 agrees well with the non-evolving
$\sigma-T_X$ relation \citep{mushotzky:97}.  Although MS2053 is not a
visually striking cluster (see Fig.~\ref{mosaic}), its dynamical mass
and X-ray temperature are consistent with it being a massive system.

\subsection{Spatial Distribution and Substructure}

\begin{figure}
\epsscale{0.8}
\plotone{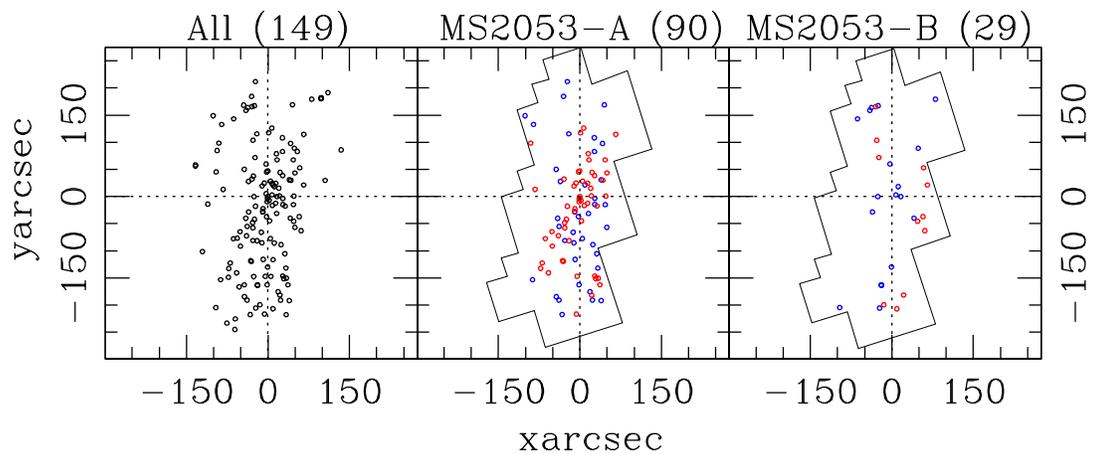}
\caption{The spatial distribution of all the spectroscopically confirmed
  cluster members are shown in the left panel.  In the middle and right
  panels, we show the members that fall on the HST/WFPC2 imaging that
  are in the main cluster (MS2053-A; middle) and in the infalling
  component (MS2053-B; left); the outline of the HST/WFPC2 mosaic is
  included in these two panels. A dynamical analysis shows that the two
  structures are gravitationally bound to each other.  The cluster
  members are separated into emission ([OII]$\lambda3727\geq5$\AA; blue
  circles) and absorption ([OII]$\lambda3727<5$\AA; red circles) line
  galaxies.  Note the extended spatial distribution and lack of
  spectral segregation in MS2053-B.
\label{xy}}
\end{figure}

Having established the bimodality in the redshift distribution of
MS2053, we now investigate whether the spatial distributions of members
in the two components also show a high degree of substructure.  The
spectroscopically confirmed cluster members have an elongated spatial
distribution (Fig.~\ref{xy}).  This is partly due to the layout of the
HST/WFPC2 mosaic which was designed to reflect MS2053's elongated
distribution first observed in the Keck imaging.  However, MS2053's
spatial distribution is mainly due to a real elongation in the
distribution of cluster members.

In MS2053-A, the absorption line members are more spatially
concentrated than the emission line members (Fig.~\ref{xy}).  However,
the spatial distributions of absorption and emission line galaxies in
MS2053-B are indistinguishable from each other, and the galaxies in
MS2053-B tend to lie at larger radii than galaxies in MS2053-A
($<R_{BCG}>=600\pm40$\hi kpc vs.  $<R_{BCG}>=380\pm40$\hi kpc).
MS2053-B's extended spatial distribution and lack of spectral
segregation is consistent with a scenario where it is in the initial
stages of being accreted by the main cluster (MS2053-A).

A useful measure of the degree of substructure in a cluster is the
Dressler-Schectman test \citep{dressler:88}.  By using redshifts and
spatial positions, the D-S test quantifies how much the local mean
redshift and velocity dispersion (as defined by the ten nearest
neighbors to each galaxy) deviate from the cluster's global values.
Considering all cluster members (149) as well as only galaxies in
MS2053-A, we find the degree of substructure in MS2053 to be
significant at the $>95$\% confidence level in both cases.  Although
MS2053 is rich, its significant substructure and bimodal redshift
distribution indicate it is a dynamically young system.


\section{Cluster Early-type Population}

\begin{figure}
\epsscale{0.5}
\plotone{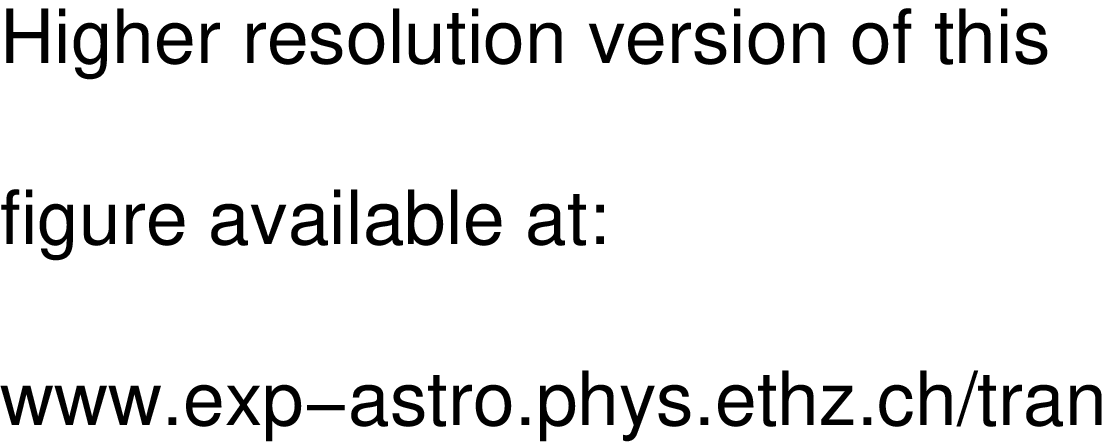}
\caption{Images of the 89 spectroscopically confirmed cluster members
  brighter than our adopted magnitude limit ($M_{Be}\leq-18$\logh) that
  fall on the HST/WFPC2 mosaic.  The images are $5''\times5''$ and
  taken in the F814W filter.  We separate the members into those in the
  main cluster (63 galaxies; prefix ``A'') and those in the infalling
  structure (26; prefix ``B'').  For each member, we include their
  morphological type, absolute $B_{e}$ magnitude, and $(B-V)_z$ color.
  We consider the following morphological types in our analysis: E/S0
  ($T\leq-1$), S0/a ($T=0$), Sa ($T=1$), spirals ($T\geq2$), and
  mergers ($T=99$).  Note the lack of S0 galaxies ($T=-2$) in MS2053.
\label{tnails}}
\end{figure}

In the following, we consider only the spectroscopically confirmed
cluster members that 1) belong to MS2053-A or MS2053-B, 2) are brighter
than $M_{Be}=-18$\logh, and 3) are on the WFPC2 mosaic.  Our
conservative magnitude limit corresponds to a spectroscopic
completeness of $\sim70$\% in the WFPC imaging (see Fig.~\ref{rates}),
and includes virtually all spectroscopically confirmed cluster members
that were visually classified.  These selection criteria reduce our
cluster sample to 63 members in MS2053-A and 26 in MS2053-B; these
galaxies are shown in Fig.~\ref{tnails}.

\subsection{Early-type Fraction}

To estimate the early-type fraction in MS2053, we follow
\citet{vandokkum:98a} and combine the E/S0's with half of the S0/a
members.  Considering only members that have been morphologically
typed, the early-type fraction in MS2053 is $50\pm8$\%.  This value is
consistent with the expected value from the observed trend in clusters
of decreasing early-type fraction with increasing redshift
\citep{dressler:97,vandokkum:00}.  It is also consistent with galaxy
clusters having a substantially higher early-type fraction than the
field: the early-type fraction in MS2053 is nearly twice that of the
field at comparable redshifts (see Table~\ref{galpops}).  However,
MS2053's higher early-type fraction is primarily due to the
main cluster; the early-type fraction in MS2053-B is actually
comparable to that of the field.

MS2053 has an unusual morphological mix in that it has no members that
are visually classified as S0 galaxies (Fig.~\ref{tnails}).  This may
be partially due to the difficulty of separating ellipticals from S0's
\citep{fabricant:00}, and it could be argued that we have missed the
S0's in MS2053 due to, e.g. inconsistencies between classifiers.
However, the same classifiers (D. Fabricant, M. Franx, and P. van
Dokkum) also morphologically typed galaxies in CL~1358+62
\citep[$z=0.33$;][]{fabricant:00}.  S0 galaxies were identified in
CL1358 whereas the same classifiers find none in MS2053.  Therefore,
the deficit of S0's in MS2053 is unlikely to be the result of
inconsistencies in the classification.

\subsection{Color-Magnitude Relation}

\begin{figure}
\epsscale{0.8}
\plotone{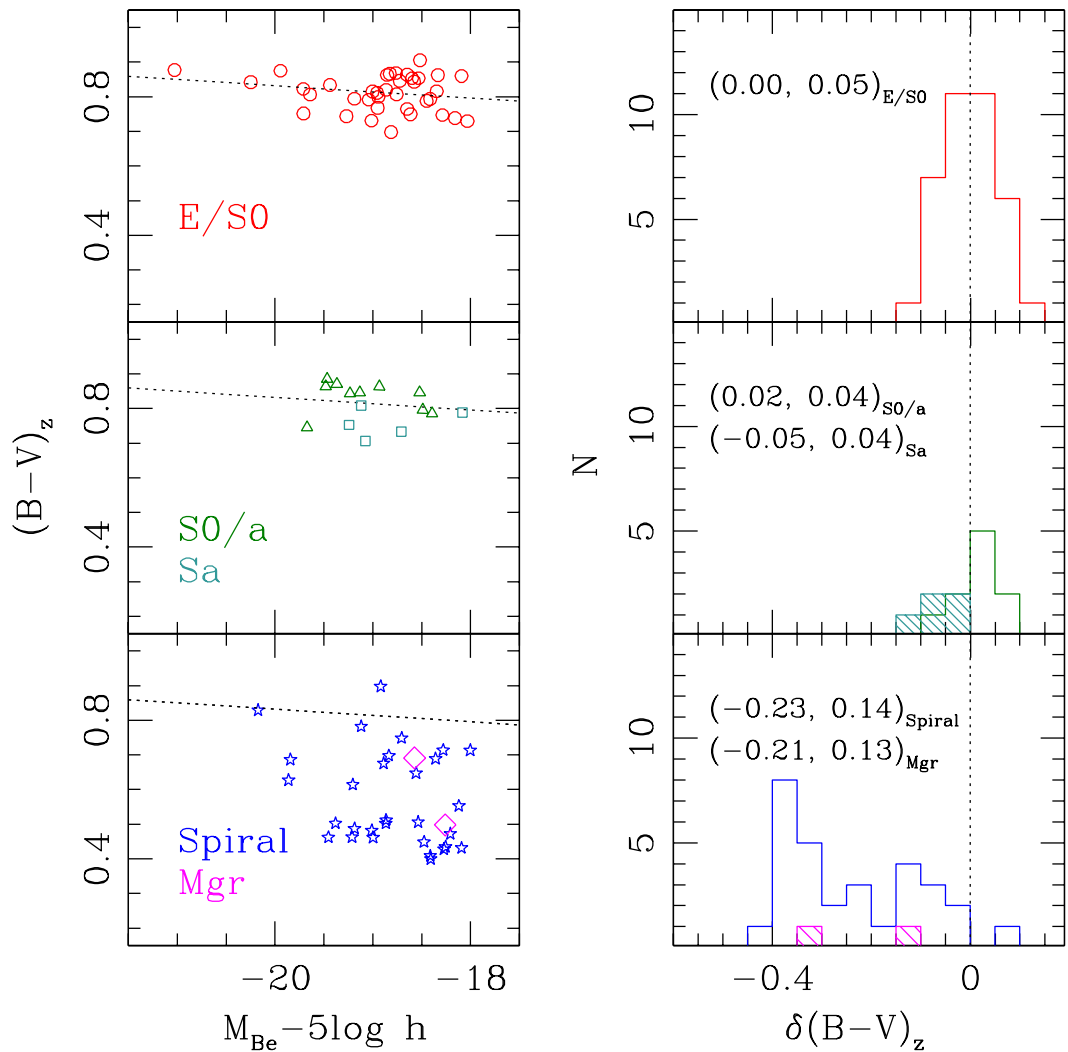}
\caption{The E/S0 galaxies in MS2053 define a red sequence that is
  well fit by the color-magnitude (CM) relation from CL1358
  \citep[$z=0.33$;][]{vandokkum:98a}.  {\it Left Panels:} The CM
  diagram for cluster members ($M_{Be}\leq-18$\logh); here we consider
  both the main cluster and group together.  The cluster galaxies are
  separated into E/S0's (small open circles), early-type spirals
  (S0/a=open triangles and Sa=open squares), and spirals and mergers
  (open stars and open diamonds).  The dotted line in each panel is the
  CM relation defined using the slope from CL1358 and normalized to the
  E/S0 members in MS2053.  {\it Right Panels:} The residuals from the
  fitted CM relation for the same galaxies.  The average offset from
  the CM relation and the scatter (RMS) are listed for the selected
  morphological types.  There is a steady trend of increasing (blue)
  offset and scatter with type, i.e.  later type cluster members tend
  to be bluer.
\label{cmd}}
\end{figure}

The color-magnitude (CM) distribution for cluster members is shown in
Fig.~\ref{cmd} where galaxies are separated by morphological type into
early-types (E/S0), early-type spirals (S0/a-Sa), spirals (Sa/b and
later), and mergers.  We adopt the slope of the CM relation measured
from CL1358 \citep[$z=0.33$;][]{vandokkum:98a} and normalize it to the
E/S0 members.  The residuals of the CM relation also are included in
Fig.~\ref{cmd}.

The early-type galaxies in MS2053 define a red sequence that is well
fit by the CM relation from CL1358.  We find no suggestion of evolution
in the slope of the CM relation, consistent with results on the CM
relation in MS~1054--03 \citep[$z=0.83$;][]{vandokkum:00}.  The observed 
scatter (RMS) of the E/S0's is 0.05; this value is larger than observed in
CL1358 \citep[$\sim0.03$;][]{vandokkum:98a}.  The larger scatter
associated with E/S0's in MS2053 may indicate that the faint
early-types are younger than the more luminous E/S0's
\citep[see][]{wuyts:04}.  We also find a trend between color and
morphological type, $i.e.$ later type members tend to be bluer, similar
to that observed at lower redshifts.

\section{Infall and the Butcher-Oemler Effect}

A recent study of A851 \citep[$z=0.41$;][]{kodama:01b} finds that the
transition from predominantly blue to red galaxy colors occurs most
often in the subclumps outside the cluster core.  However, a link
between galaxy infall and a high fraction of blue cluster galaxies has
yet to be conclusively established.  Here we demonstrate that: 1) the
two structures making up MS2053 are gravitationally bound to each other
and will eventually merge; 2) MS2053's blue fraction is elevated by the
galaxies in the infalling structure; and 3) the infalling galaxies as a
whole are indistinguishable from those in the field.

As in the previous section, we consider only spectroscopically
confirmed members brighter than $M_{Be}=-18$\logh~that fall on the
HST/WFPC2 mosaic (Table~\ref{galpops}; Fig.~\ref{tnails}).

\subsection{Evidence of Infalling Galaxies}

To confirm that MS2053-A and MS2053-B are gravitationally bound to each
other and will eventually merge into one system, we follow
\citet{beers:82} and treat the system as a two-body problem.  From
their Eq. 15, a bound system has

\begin{equation}
V_r^2 R_p \leq 2GM \sin^2(\alpha) \cos(\alpha) 
\end{equation}

\noindent where $V_r$ is the relative velocity, $R_p$ the projected
separation, $M$ the total mass of the system, and $\alpha$ the
projection angle with respect to the plane of the sky.  

MS2053-A and MS2053-B are virtually superimposed on the sky (see
Fig.~\ref{xy}): the difference in the average projected radius ($R_p$)
of the galaxies in the two structures is 97\hi kpc.  The total mass of
the system is $1.2\times10^{15}M_{\odot}$ and the difference in the
mean redshifts of MS2053-A and MS2053-B gives $V_r=2700$\kms.  Because
the projection angle $\alpha$ is the only unknown parameter, we adopt a
simple approach and determine whether the two structures are bound for
$0^{\circ}\leq \alpha\leq90^{\circ}$.  We find that as long as
$15^{\circ}<\alpha<86^{\circ}$, MS2053-A and MS2053-B are bound.  Thus
given a random distribution of projection angles, the probability that
the two structures are bound is $\sim80$\%, $i.e.$ very likely.

As another check, we compare the relative velocity of the two
structures ($V_r$) to the escape velocity ($V_{esc}$) of the main
cluster (MS2053-A).  Using $M_V(A)=1.1\times10^{15} M_{\odot}$ and
$R_V(A)=1.07$\hi~Mpc (Table~\ref{stats}), we estimate
$V_{esc}\sim3000$\kms.  Since $V_{r}<V_{esc}$, the two structures are
bound.  In addition, the similarity in the two velocities suggests that
MS2053-B is infalling.

An infalling scenario is supported by the properties of the galaxies in
MS2053-B.  The fraction of star-forming galaxies in MS2053-B is
comparable to that of our field sample (67\% vs. 61\%), and it is
$\sim2$ times larger than in MS2053-A (see Table~\ref{galpops}).  This
is also the case for the blue galaxies.  Since star formation in
infalling galaxies is effectively quenched within a few Gyr
\citep{balogh:00}, $i.e.$ a core-crossing time, their activity
indicates the galaxies in MS2053-B have yet to pass through the core of
the main cluster.

\subsection{Fraction of Blue Cluster Members}

Using the classical definition of a blue cluster galaxy \citep[$\Delta
(B-V)_z\leq-0.2$;][]{butcher:78}, we measure the fraction of blue
members in MS2053 to be $24\pm5$\% (see Table~\ref{galpops}).
MS2053's fraction of blue cluster members agrees with the expected
value from the original trend found by \citet{butcher:84} of
increasing blue fraction with redshift.  It is also consistent with
the correlation between blue fraction and redshift measured by
\citet{ellingson:01} using the CNOC1 cluster sample ($0.18<z<0.55$).

We have defined MS2053's blue cluster fraction using our sample of
spectroscopically confirmed cluster members ($M_{Be}\leq-18$\logh)
rather than using the classic photometric approach \citep{butcher:84}.
The advantage of using a spectroscopic sample to determine MS2053's
blue fraction is that we are free from the usual uncertainty due to
field contamination that is associated with photometric studies.
However, we note for comparison that our blue fraction is very similar
to that determined photometrically
\citep[$f_b=25\pm4$\%;][]{fairley:02}.

\subsection{Origin of Blue Cluster Members}

We find that MS2053's high fraction of blue cluster members is due
primarily to the galaxies in MS2053-B and not the main cluster: more
than half of the blue galaxies are associated with MS2053-B
(Fig.~\ref{zhist}).  The fraction of blue galaxies in the main cluster
is only 13\% and is comparable to that of clusters at $z\leq0.2$
\citep{butcher:84,ellingson:01}.  However, the fraction of blue
galaxies in MS2053-B is over a factor of 3 higher (46\%).

The galaxy population in MS2053-B is remarkably similiar to that of the
general field.  To quantify this, we compare the color-magnitude
distributions of both MS2053-A and MS2053-B to that of the field
(Fig.~\ref{cmd2}).  In all three environments, the early-type galaxies
follow the defined CM relation and the luminosity distributions are
indistinguishable.  However, while the fraction of blue galaxies in
MS2053-B is comparable to that in the field (56\%), both are
significantly higher than in MS2053-A.  Using the K-S test, the color
distribution of MS2053-B's galaxies is indistinguishable from that of
the field, whereas MS2053-A's galaxies differ at the $>95$\% confidence
level.  We also note that the morphological distribution of galaxies in
MS2053-B is indistinguishable from the field.

Finally, we compare the estimated internal velocity dispersions (see
\S3.5) of the galaxies in MS2053-A, MS2053-B, and the field
(Fig.~\ref{MBz_nsigma}).  Using the K-S test, we find that the
$\sigma_{est}$ distribution of MS2053-B is indistinguishable from that
of the field.  However, MS2053-A's galaxies differ from the field at
the $>95$\% confidence level.  We conclude that MS2053's elevated
fraction of blue members is due to a population of infalling galaxies
that are indistinguishable from the field.

\section{Spectroscopic Butcher-Oemler Effect}

Our spectroscopic survey not only enables us to determine the fraction
of blue galaxies in MS2053 without contamination from the field, it
also allows us to test whether the high fraction of blue members is
reflected in the spectroscopic properties of the cluster members.  By
separating the cluster members into absorption line, emission line, and
post-starburst galaxies, we measure the fraction of active galaxies in
MS2053.  To confirm that the observed [OII] emission is due to star
formation rather than emission from active galactic nuclei, we utilize
Chandra observations of the MS2053 field.  We then determine whether
the blue members in MS2053 are related to its star-forming population.

\subsection{Emission Line Members}

MS2053 has an unusually high fraction of emission line galaxies (50\%):
of the 111 cluster members for which we can measure [OII], 56 have
[OII]$\geq5$\AA.  In comparison to clusters at $z\sim0.3$ with
comparable velocity dispersions \citep{fisher:98,balogh:02}, MS2053 has
{\it more than twice} the number of active members.  As found in
earlier studies of lower redshift clusters
\citep{couch:87,dressler:92}, MS2053's high fraction of blue members is
reflected in its high fraction of spectroscopically active members.  We
find that like the blue cluster members, the fraction of emission line
galaxies in MS2053-B is significantly higher than in the main cluster
(see Table~\ref{galpops}).

\subsection{AGN Activity}

To determine whether any of MS2053's emission line galaxies harbor
active galactic nuclei (AGN), we analyse archival X-ray observations
from Chandra.  Two overlapping pointings of the MS2053 field were taken
with ACIS-I ($16.9'\times16.9'$).  Each pointing was $\sim45$~ksec,
giving an approximate flux detection limit of
$f_X[2-10keV]>10^{-15}$~erg~s$^{-1}$~cm$^{-2}$ \citep{martini:02}; this
corresponds to $L_X>7.4\times10^{41}$~erg~s$^{-1}$ at $z=0.59$ and is
deep enough to detect the majority of AGN \citep{norman:04}.

Of the 149 confirmed cluster members, we find only the brightest
cluster galaxy (BCG; \#1667) is detected in the X-ray.  However, the
BCG has no detectable [OII] emission.  Of the 56 cluster members with
[OII]$\geq5$\AA, none are X-ray detected to the given flux limit.
Thus the [OII] emission measured in MS2053's members is most likely
due to ongoing star formation rather than AGN activity.

\subsection{Ongoing Star Formation}

\begin{figure}
\epsscale{0.8}
\plotone{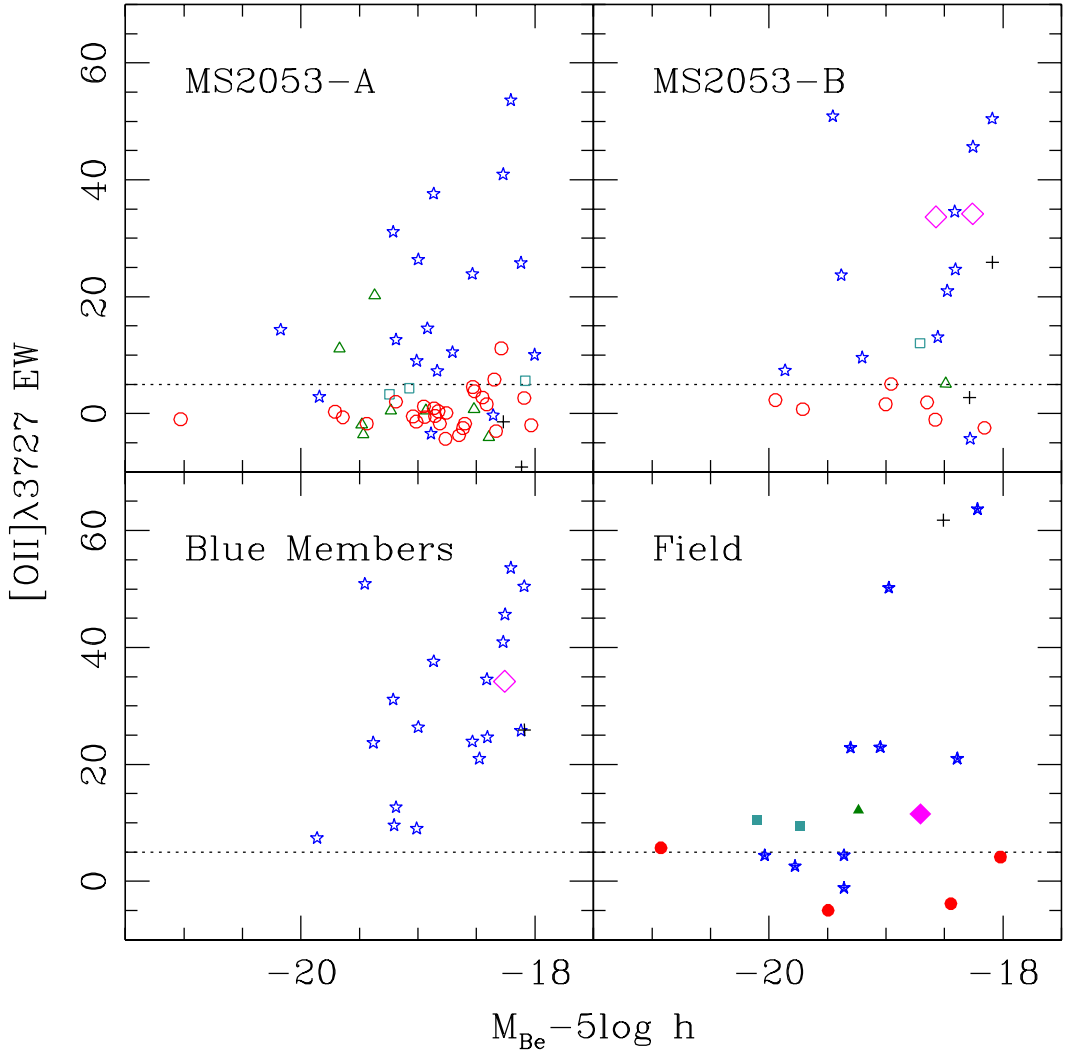}
\caption{OII$\lambda3727$ equivalent widths versus absolute
  $B_{e}$ magnitude for galaxies in MS2053-A (upper left), galaxies in
  the MS2053-B (upper right), blue cluster members from both structures
  (lower left), and the field (lower right); only galaxies brighter
  than $M_{Be}=-18$\logh~are considered.  The symbols are as in
  Fig.~\ref{cmd} and non-typed members (plus symbols) are also
  included.  The dotted horizontal line in all panels denotes our
  cut-off for emission line ([OII]$\geq5$\AA) systems.  
  We find a trend in the blue cluster galaxies of
  increasing [OII] EW with decreasing luminosity; this is consistent
  with fainter members having younger luminosity-weighted ages.
  In addition, the [OII] EW
  distribution of MS2053-B is indistinguishable from that of the field,
  but MS2053-A's [OII] EW distribution differs from the field at the
  $>95$\% level.  This indicates that galaxies in the infalling 
  structure and the field share a common parent population.
\label{MBz_OII}}
\end{figure}

In Fig.~\ref{MBz_OII}, we show the distribution of [OII] equivalent
width versus absolute $B_{e}$ magnitude as a function of morphological
type for both the cluster members and field sample; we consider only
galaxies with $M_{Be}\leq-18$\logh~that have HST/WFPC2 imaging.  Not
surprisingly, the majority of early-type galaxies in both environments
are absorption line ([OII]$<5$\AA) systems while the majority of the
late-types are emission line ([OII]$\geq5$\AA) galaxies.  Note the
trend in the blue galaxies of increasing [OII]  with decreasing
luminosity; this trend is signficant at the $>95$\% level using the
Spearman rank test \citep{press:92}.

One striking result from Fig.~\ref{MBz_OII} is that {\it all} of the
blue galaxies in MS2053 are emission line systems (Fig.~\ref{MBz_OII}),
$i.e.$ forming stars, and virtually all of them are spirals.  However,
not all of the emission line members are blue.  In fact, there are 15
star-forming members that are red, and they include early-types and
S0/a's; these may be the dusty, star-forming members discussed in
\citet{poggianti:99}.  Although blue colors do identify star-forming
members, using colors alone underestimates the total star formation
activity in the cluster.

\subsection{Infalling Galaxies vs. the Field}

Results from several studies suggest that infall does not induce excess
star formation in the infalling galaxies relative to the field
\citep{balogh:97,ellingson:01,kodama:01a}.  To test if this is the case
in MS2053, we compare the [OII] and Balmer EW distributions of galaxies
in MS2053-B to those in the field.  We consider only the galaxies in
MS2053-B because we are relatively confident that they are falling into
the cluster for the first time, whereas this is not the case if we
include galaxies from MS2053-A.  To enable comparison to \S6 and \S7,
we adopt our usual magnitude limit ($M_{Be}\leq-18$\logh) and
requirement of WFPC2 imaging.

In terms of their [OII] and Balmer EW's, a K-S test finds MS2053-B's
galaxies share a common parent population with the field galaxies.
Assuming [OII] and Balmer EW traces star formation in the same manner
in both environments, we find no indication of enhanced activity in
MS2053-B's galaxies relative to the field.  However, we note that we
cannot say whether star formation is enhanced later as MS2053-B and
MS2053-A continue to merge.

As shown here, the similarities between the infalling galaxies and
those in the field are not limited to their photometric properties (see
\S7.3).  Their [OII] and Balmer EW distributions are indistinguishable,
and their average [OII] EW are the same: $16.6\pm3.6$\AA~and
$16.5\pm5.0$\AA~for MS2053-B and the field respectively.  In
comparison, the average [OII] EW of MS2053-A is $\sim2.5$ times lower
($6.2\pm1.6$\AA), and its [OII] EW distribution differs from that of
the field at the $>95$\% confidence level.  More than half of MS2053's
star-forming members belong to the infalling structure (MS2053-B), and
the infalling galaxies are spectroscopically indistinguishable from
that of the field population.

\section{Descendants of Blue Cluster Galaxies}

Having established that MS2053's elevated fraction of blue/star-forming
galaxies is due primarily to an infalling structure, we now address
what the descendants of these blue galaxies can be.  Are these new
members the progenitors of the S0 galaxies found in nearby clusters?
In the following discussion, we consider blue galaxies from MS2053-A
and MS2053-B together, and adopt our usual magnitude limit
($M_{Be}\leq-18$\logh) and requirement of WFPC2 imaging.

\begin{figure}
\epsscale{0.8}
\plotone{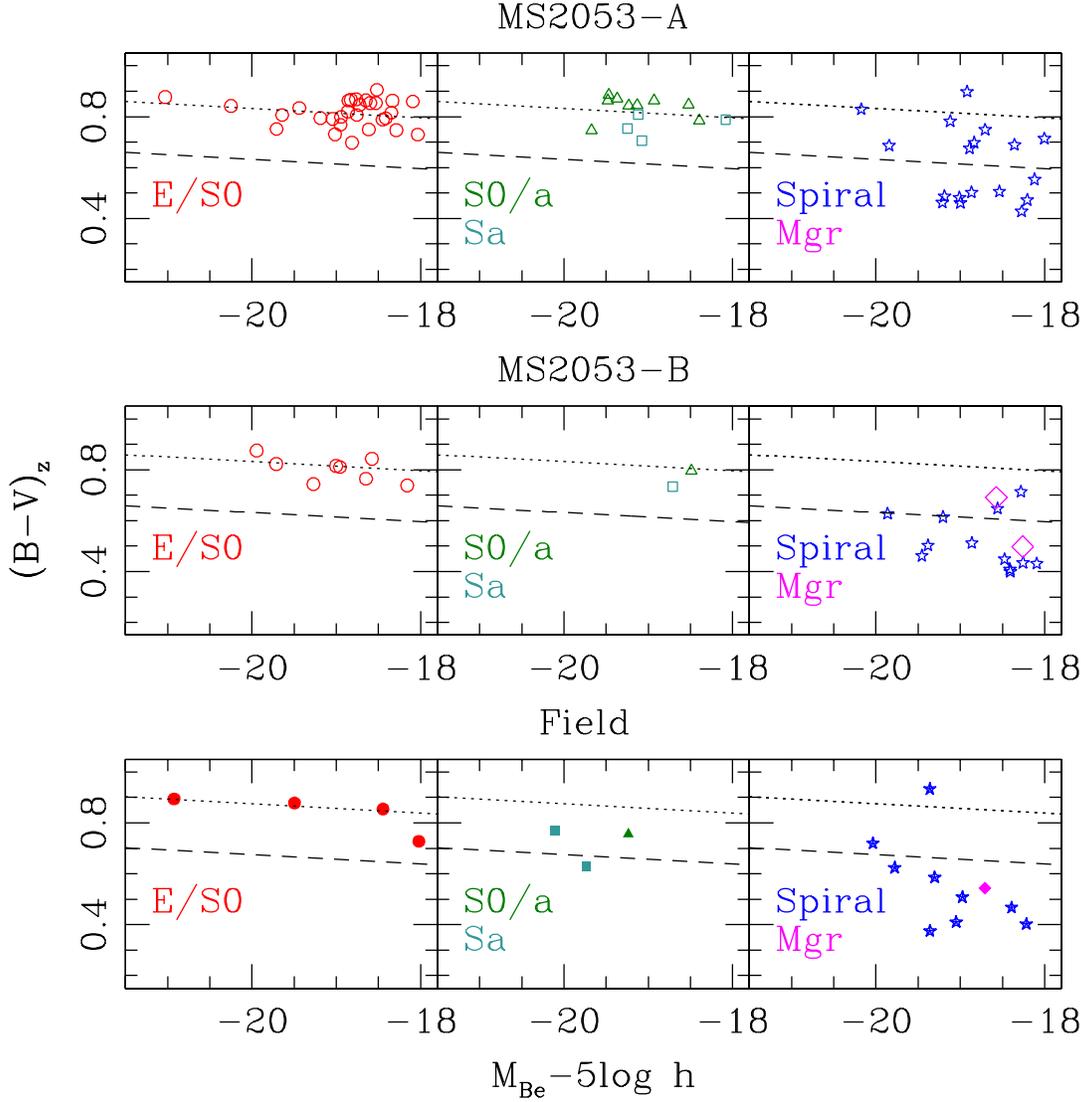}
\caption{Color-magnitude diagrams for galaxies in the main cluster
  (MS2053-A; top), the infalling component (MS2053-B; middle), and the
  field (bottom).  We denote the classic definition of a blue galaxy as
  one having $\Delta(B-V)_z\leq-0.2$ \citep[dashed
  line;][]{butcher:84}; the symbols and dotted line are as in
  Fig.~\ref{cmd}.  Using the K-S test, we find that the color and
  luminosity distributions of the galaxies in MS2053-B are
  indistinguishable from those of our field sample.  However, the
  colors of the galaxies in the main cluster (MS2053-A) differ from
  those of the field at the $>95$\% confidence level.  MS2053-A also
  has a number of red spirals.
\label{cmd2}}
\end{figure}

Similar to what has been observed in other intermediate redshift
clusters \citep{couch:98}, we find MS2053's blue galaxies tend to be
fainter than $L^{\ast}$.  None are brighter than
$M_{Be}=-20$\logh~(see Fig.~\ref{cmd2}), and their average absolute
$B_{e}$ magnitude is $-18.7$\logh.  The luminosities of MS2053's blue
cluster galaxies indicate they will evolve into members with
$L<L^{\ast}$.  Assuming the blue cluster galaxies evolve passively with
time, they will fade and redden to eventually populate the faint end of
the CM relation, as suggested by \citet{kodama:01a}.

\begin{figure}
\epsscale{0.8}
\plotone{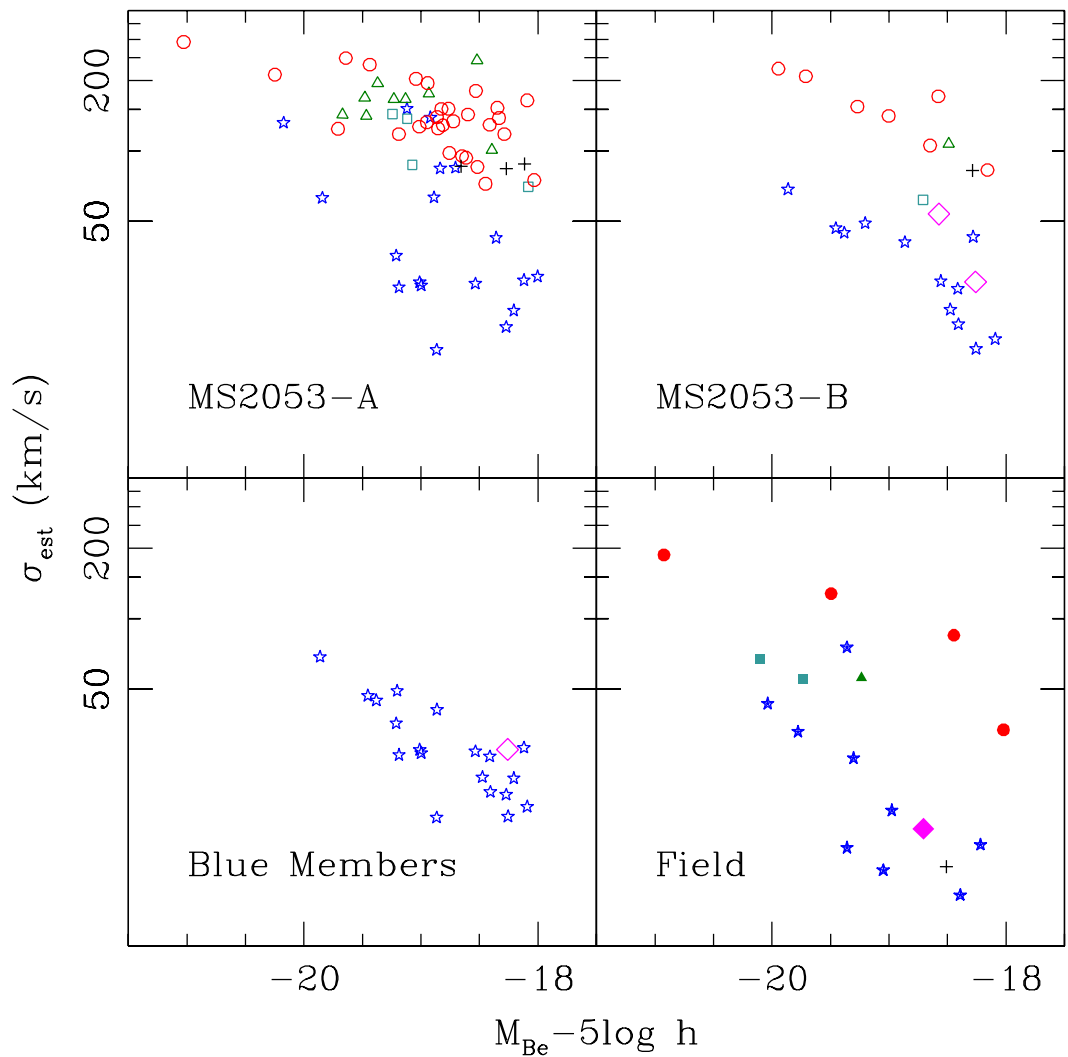}
\caption{We show estimated internal velocity dispersions versus
  absolute $B_{e}$ magnitude for galaxies in MS2053-A (top left) and
  MS2053-B (top right), blue cluster galaxies from both structures
  (bottom left), and the field sample (bottom right).  We consider only
  galaxies brighter than $M_{Be}\leq-18$\logh, and the symbols are as
  in Fig.~\ref{cmd}.  All of the blue galaxies in MS2053 are fainter
  than $M_{Be}=-20$\logh~and have $\sigma_{est}<100$\kms; they can only
  evolve into faint, low mass members.  In addition, the galaxies in
  MS2053-B share a common parent $\sigma_{est}$ distribution with those
  in the field, while those in MS2053-A differ at the $>95$\% CL.
  MS2053's elevated fraction of blue members is due to an infalling
  structure whose galaxies are indistinguishable from the field
  population.
\label{MBz_nsigma}}
\end{figure}

The estimated internal velocity dispersions ($\sigma_{est}$) of the
blue cluster galaxies also indicates they will remain low mass systems.
With only one exception, all of the blue galaxies in MS2053 have
$\sigma_{est}<50$\kms~(see Fig.~\ref{MBz_nsigma}); these blue galaxies
cannot evolve into massive cluster members.  Their low $\sigma_{est}$
are consistent with \citet{depropris:03}'s suggestion that the B-O
effect is largely due to low mass, star-forming galaxies.

MS2053 has an unusual morphological mix of galaxies
compared to clusters at $z\sim0$: it has no S0's but it has twice the
number of blue spirals compared to clusters at lower redshifts.  Note
that a deficit of S0's has also been observed in other clusters at
intermediate redshifts \citep{dressler:97,lubin:02}.  Assuming MS2053
will eventually have a galaxy population similar to that of equally
massive clusters at lower redshifts, $e.g.$ CL1358 or Coma, this means
that MS2053 must develop a sizeable population of S0 galaxies within
the next $\sim5$ Gyr.  Given its high fraction of blue spirals, the
obvious solution is to transform these late-types into the missing S0
population.

The conversion in MS2053 of the blue cluster galaxies into faint, low
mass S0's would be consistent with the luminosity-weighted ages of low
mass S0 galaxies in Coma \citep{poggianti:01b}.  Although the
ellipticals in Coma have been quiescent over the last $\sim5$ Gyr, more
than 40\% of the S0's have formed stars during the same period and so
have younger luminosity-weighted ages.  In addition, the fraction of
S0's in Coma with recent activity increases with decreasing luminosity.
This latter point is consistent with our observed trend in MS2053's
blue galaxies of increasing [OII] EW with decreasing luminosity, $i.e.$
lower mass blue cluster members have younger luminosity-weighted ages.

Our analysis shows that the blue cluster galaxies can be the missing
link between the deficit of S0's observed at intermediate redshifts and
their relative abundance in the local universe.  Not only does MS2053
have an overabundance of blue, low mass spirals, their spectral
properties are consistent with the extended star formation histories
observed in low mass cluster S0's at lower redshifts.  Although they
originated in the field, these blue galaxies are the likely progenitors
of many of the $L<L^{\ast}$ S0 galaxies seen in nearby clusters.

Finally, our results highlight a serious problem that still remains.
Although we have established the link between blue cluster galaxies and
{\it low mass} S0 members, our analysis does not account for the many
bright ($L>L^{\ast}$) S0 galaxies in nearby clusters.  This may be
partially due to the difficulty of separating S0's from ellipticals at
higher redshift, $i.e.$ some of MS2053's S0 galaxies may have been
visually classified as ellipticals.  Evolutionary effects can also
increase the number of luminous S0's in MS2053.  For example, there are
several bright, red, massive spirals in MS2053 (see Figs.~\ref{cmd2} \&
\ref{MBz_nsigma}) that are viable progenitors of the luminous S0's at
lower redshift.  We have also identified several bright, red, massive,
passive spirals in MS~1054--03 \citep[$z=0.83$;][]{tran:03b} that may
be the progenitors of massive cluster S0's at $z\sim0$. However, we
stress that our current analysis does not address the origin of the
bright S0 galaxies in nearby clusters.

\section{Conclusions}

We have studied the galaxy populations in MS2053 using high resolution,
wide-field imaging from HST/WFPC2 and an extensive spectroscopic survey
completed with Keck/LRIS.  MS2053 is an X-ray lumininous cluster at
$z=0.5866\pm0.0011$ with a bimodal redshift distribution.  It is
composed of a main cluster (MS2053-A) and an infalling structure
(MS2053-B) that are gravitationally bound to each other.  Of the 149
spectroscopically confirmed cluster members, 113 belong to MS2053-A and
36 to MS2053-B.  The velocity dispersions of MS2053-A and MS2053-B are
$865\pm71$\kms~and $282\pm51$\kms, and the total dynamical mass of the
system is $1.2\times10^{15}M_{\odot}$.  MS2053 also lenses a background
galaxy at $z=3.1462$.

MS2053 is a classic Butcher-Oemler cluster: 24\% of its members are
blue galaxies, and an even higher fraction (44\%) are star-forming.
However, more than half of the blue/star-forming members belong to the
infalling structure (MS2053-B).  Unlike previous studies that have
found only indirect evidence for the link between the Butcher-Oemler
effect and galaxy infall, our unique dataset enables us to show
conclusively that this is the case in MS2053.

Comparing MS2053's infalling galaxies to a field sample at
approximately the same redshift, we find that they share a common 
parent population.  Galaxies in MS2053-B span the same ranges
in luminosity, color, estimated velocity dispersion ($\propto$ mass),
and [OII]$\lambda3727$ equivalent width as those in the field.  Not 
only is MS2053's fraction
of blue/star-forming members boosted due to an infalling component, the
galaxies in MS2053-B are indistinguishable from that of the field
population.  We do not find any evidence of enhanced star formation in
the infalling galaxies relative to the field, although we note this may
change as MS2053-A and MS2053-B continue to merge.

Given that MS2053 currently has no S0 galaxies but rather an
overabundance of blue, star-forming, low mass late-types, it is very
likely that many of these spirals will evolve into low mass S0
galaxies.  Such a scenario is consistent with the extended star
formation histories of low mass S0's in lower redshift clusters.  If
MS2053's blue galaxies do evolve into S0's, their luminosities and
estimated internal velocity dispersions indicate they will be faint
($L<L^{\ast}$), low mass members.

Taken as a whole, our observations suggest that most of the blue
members in MS2053, and ultimately most of its low mass S0's, originate
in the field.

\acknowledgments

The authors thank D. Fabricant for help with the visual classifications
and D. Magee with the cluster redshift survey.  K. Tran thanks M.
Santos and J. Diemand for useful discussions.  K. Tran gratefully
acknowledges support from the Swiss National Science Foundation. A.
Gonzalez is supported by an NSF Astronomy and Astrophysics Postdoctoral
Fellowship under award AST-0407085. Additional support from NASA HST
grants GO-06745.01, GO-07372.01, and GO-08220.03, and NASA grant
NAG5-7697 also are acknowledged.  The authors thank the entire staff of
the W. M.  Keck Observatory for their support, and extend special
thanks to those of Hawaiian ancestry on whose sacred mountain we are
privileged to be guests.

\bibliographystyle{/home/vy/aastex/apj}
\bibliography{/home/vy/aastex/tran.bib}


\clearpage

\clearpage

\clearpage

\clearpage

\clearpage

\clearpage

\clearpage

\clearpage

\clearpage

\clearpage

\clearpage

\begin{deluxetable}{lrrrrrrr}
\tablecolumns{7}
\tablewidth{0pc}
\tablecaption{Cluster Kinematics\label{stats}}
\tablehead{
\colhead{Group} & \colhead{Number}  & \colhead{$z$-range} & 
\colhead{$\bar{z}$} &\colhead{$\sigma_{1D}$ (km~s$^{-1}$)} & 
\colhead{$R_{V}$ (Mpc)\tablenotemark{b} } & 
\colhead{$M_V$ ($M_{\odot}$)\tablenotemark{b}}}
\startdata
MS2053-All\tablenotemark{a} & 149 & $0.57\leq z\leq0.605$ & $0.5866\pm0.0011$&
    $1523\pm95$ &1.15  & $3.7\times10^{15}$ \\
MS2053-A& 113 & $0.57\leq z<0.595$& $0.5840\pm0.0005$& $865\pm71$
                &1.07 & $1.1\times10^{15}$ \\
MS2053-B&  36 & $0.595\leq z\leq0.605$& $0.5982\pm0.0003$& $282\pm51$
                &1.13 & $1.1\times10^{14}$ \\
Field &  38 & $0.54\leq z\leq 0.64$ & $0.58\pm0.01$ & \nodata & \nodata
& \nodata \\
\enddata
\tablenotetext{a}{Note that when considering all cluster members, the
  velocity dispersion and virial mass are both significantly
  overestimated due to MS2053's bimodal redshift distribution.}
\tablenotetext{b}{Virial radii and masses determined using
  $\Omega_M=0.3, \Omega_{\Lambda}=0.7$, and $H_0=100h$\kms~Mpc$^{-1}$
  cosmology.} 
\end{deluxetable}

\begin{deluxetable}{lrrrr}
\tablecolumns{5}
\tablewidth{0pc}
\tablecaption{Galaxy Populations\tablenotemark{a}\label{galpops}}
\tablehead{
\colhead{Property} & \colhead{MS2053-All} & \colhead{MS2053-A} 
&\colhead{MS2053-B} & \colhead{Field} }
\startdata
Number\tablenotemark{b}& 89 (84)  & 63 (60)  & 26 (24) & 18 (17) \\
E/S0               & 37       & 29       &  8      &  4 \\
S0/a               & 10       &  9       &  1      &  1 \\
Sa                 &  5       &  4       &  1      &  1 \\
Spiral             & 30       & 18       & 12      &  9 \\
Merger             &  2       &  0       &  2      &  1 \\
Early-Types \tablenotemark{c}& 50\%     & 56\%     & 35\%    & 26\%\\
Late-Types \tablenotemark{c} & 50\%     & 44\%     & 65\%    & 74\%\\
Red  [$\Delta(B-V)_z>-0.2$]   & 76\%  & 87\% & 54\%  & 44\% \\
Blue [$\Delta(B-V)_z\leq-0.2$]& 24\%  & 13\% & 46\%  & 56\% \\
Absorption Line\tablenotemark{d}& 51\%  & 60\% & 29\%  & 39\% \\
Emission Line\tablenotemark{d}  & 44\%  & 34\% & 67\%  & 61\% \\
E+A\tablenotemark{d}            & 5\%   &  5\% &  4\%  &  0\% \\
\enddata
\tablenotetext{a}{Considering only the galaxies that fall on the
  HST/WFPC2 mosaic that are brighter than our adopted magnitude limit
  of $M_{Be}\leq-18$\logh.}
\tablenotetext{b}{The total number of galaxies where the number in
  parentheses corresponds to the number that are visually typed into
  Hubble classes.}
\tablenotetext{c}{The early-type fraction is determined by combining
  all the E/S0 members ($T\leq-1$) with half of the S0/a members
  ($T=0$).  The late-type fraction is all members with $T\geq1$
  combined with half of the S0/a's.}
\tablenotetext{d}{Spectral types are defined as emission line
  ([OII]$\lambda3727\geq5$\AA), passive ([OII]$<5$\AA), and E+A
  ((H$\delta+$H$\gamma)/2\geq4$\AA, [OII]$<5$).}
\end{deluxetable}

\end{document}